Main Manuscript for

# An Isolated Water Droplet in the Aqueous Solution of a Supramolecular Tetrahedral Cage

Federico Sebastiani,[1] Trandon A. Bender,[2] Simone Pezzotti[1], Wan-Lu Li[2,3], Gerhard Schwaab,[1] Robert G. Bergman,[2] Kenneth N. Raymond,[2] F. Dean Toste,[2] Teresa Head-Gordon[2,3], Martina Havenith[1]

1 Lehrstuhl für Physikalische Chemie II, Ruhr-Universität Bochum, 44780 Bochum, Germany
2 Chemical Sciences Division, Lawrence Berkeley National Laboratory, and Department of Chemistry, University of California, Berkeley, California 94720-1460
3 Pitzer Center for Theoretical Chemistry, University of California, Berkeley, California 94720-1460

*Corresponding authors: Robert G. Bergman, Kenneth N. Raymond, F. Dean Toste, Teresa Head-Gordon, Martina Havenith

**Email:**
*rbergman@berkeley.edu
*raymond@socrates.berkeley.edu
*fdtoste@berkeley.edu
*martina.havenith@rub.de
*thg@berkeley.edu

**Abstract**

Water under nanoconfinement at ambient conditions has exhibited low-dimensional ice formation and liquid-solid phase transitions, but with structural and dynamical signatures which map onto known regions of waters phase diagram. Using THz absorption spectroscopy and ab initio molecular dynamics, we have investigated the ambient water confined in a supramolecular tetrahedral assembly, and determined that a distinct network of 9±1 water molecules is present within the nanocavity of the host. The low-frequency absorption spectrum and theoretical analysis of the water in the $Ga_4L_6^{12-}$ host demonstrate that the structure and dynamics of the encapsulated droplet is distinct from any known phase of water. A further inference is that the release of the highly unusual encapsulated water droplet creates a strong thermodynamic driver for the high affinity binding of guests in aqueous solution for the $Ga_4L_6^{12-}$ supramolecular construct.

**Significance Statement**

The chemical and physical versatility of ambient liquid water continues to offer scientific surprises. Using a supramolecular assembly that catalyzes substrates in water, we show that the nanocage encapsulates a water cluster that is structurally and dynamically distinct from any known phase of water that helps explain its catalytic activity.

**Main Text**

**INTRODUCTION**

Supramolecular capsules create internal cavities that are thought to act like enzyme active sites.[1] As aqueous enzymes provide inspiration for the design of supramolecular catalysts, one of the goals of supramolecular chemistry is the creation of synthetic "receptors" that have both a high affinity and a high selectivity for the binding of guests in water.[2-3] The $Ga_4L_6^{12-}$ tetrahedral assembly formulated by Raymond and co-workers represents an excellent example of a water-soluble supramolecular cage that has provided host interactions that promotes guest encapsulation. Using steric interactions and electrostatic charge to chemically position the substrate while shielding the reaction from solvent, this host has been shown to provide enhanced reaction rates that approach the performance of natural biocatalysts.[4-10] Moreover, aqueous solvation of the substrate, host, and encapsulated solvent also play an important role in the whole catalytic cycle. In particular, the driving forces that release water from the nanocage host to favour the direct binding with the substrate is thought to be a critical factor in successful catalysis, but is challenging to probe directly.[7-8, 11-13, 14]

In both natural and artificial nanometer-sized environments, confined water displays uniquely modified structure and dynamics with respect to the bulk liquid.[15-18] Recently, these modified properties were also found to have significant implications for the mechanism and energetics of reactions taking place in confined water with respect to those observed in bulk aqueous solution.[19-21] In a pioneering study on supramolecular assemblies, Cram and collaborators concluded that the interior of those cages is a "new and unique phase of matter" for the incarcerated guests.[22] In more recent studies it was postulated that, similar to graphitic and zeolite nanopores[23-24], confined water within supramolecular host cavities is organized in stable small clusters (($H_2O)_n$, with n=8-19) which are different from gas phase water clusters.[25] In these studies, the hydrogen-bonded water clusters were reported to be mostly ice- or clathrate-like by X-ray and neutron diffraction in the solid state at both ambient and cryogenic temperatures.[26-32] However, to the best of our knowledge, such investigations have not characterized the $Ga_4L_6^{12-}$ supramolecular tetrahedral assembly in the liquid state near room temperature and pressure, where the $[Ga_4L_6]^{12-}$ capsule can perform catalytic reactions.[6, 8-9]

Here, we use THz absorption spectroscopy and ab initio molecular dynamics to characterize low-frequency vibrations and structural organization of water in the nanoconfined environment. THz is ideally suited to probe the intermolecular collective dynamics of the water hydrogen bond network with extremely high sensitivity, as illustrated for different phases of water[33-38], and for aqueous solutions of salts, osmolytes, alcohols and amino acids.[36, 39-42] The THz spectra of the water inside the nanocage has been quantitatively reproduced with ab initio molecular dynamics (AIMD), allowing us to confidently characterize the water network in the cage in order to provide a more complete dynamical and structural picture. We have determined that the spectroscopic signature of the confined water in the nanocage is a dynamically arrested state whose structure bears none of the features of water at any alternate thermodynamic state point such as pressurized liquid or ice. Our experimental and theoretical study provides insight into the role played by encapsulated water in supramolecular catalysis, creating a low entropy and low enthalpy water droplet readily displaced by a catalytic substrate.

**RESULTS**

In the presence of strongly binding cationic salts such as tetraethylammonium salts ($[Et_4N]^+$) at a 1:1 guest:host concentration, all the cations are quantitatively encapsulated in the cavity (see Methods) given that their internal binding constant is almost three orders of magnitude larger than external binding, with minimal encapsulated water molecules present (Figure 1b).[5-6, 14, 43] At ambient conditions, the aqueous soluble naphthalene-based supramolecular host is proposed to contain some number of water molecules within the intramolecular space as well as interfacial solvent molecules associated with the external structure (Figure 1a). However, the number and the nature of the water molecules encapsulated within the cage is not known, with or without the $[Et_4N]^+$ guest.

*Experimental Results.* The THz absorption spectra from $\nu$ = 50 to 450 cm$^{-1}$ at 293K were recorded for the water-filled and for the encapsulated $[Et_4N]^+$ guest at 20 mM (Figure 1c), and a second concentration of 10 mM is reported in the Supplementary Information.

These spectra were differenced from the bulk water spectrum to remove the contribution of the background solvent, and to determine the change in absorbance as a function of frequency, $\Delta\alpha(\nu)$ of the guest-filled complexes (for details see pages S1-S3 in the Supporting Information). We find that the water-filled cavity displays an increased THz absorption with respect to that of the encapsulated salt in the 100-270 cm$^{-1}$ range, which is characteristic of the changes in the intermolecular hydrogen bond stretching of the water in and around the cage that differs from bulk water at ambient temperatures. In addition, peaks above 270 cm$^{-1}$ were observed and assigned to the intramolecular modes of the Ga$_4$-L$_6^{12-}$ tetrahedral host that overlap with the broad librational band of water. Even so, we are only interested in the confined water signatures that occur at frequencies below 270 cm$^{-1}$.

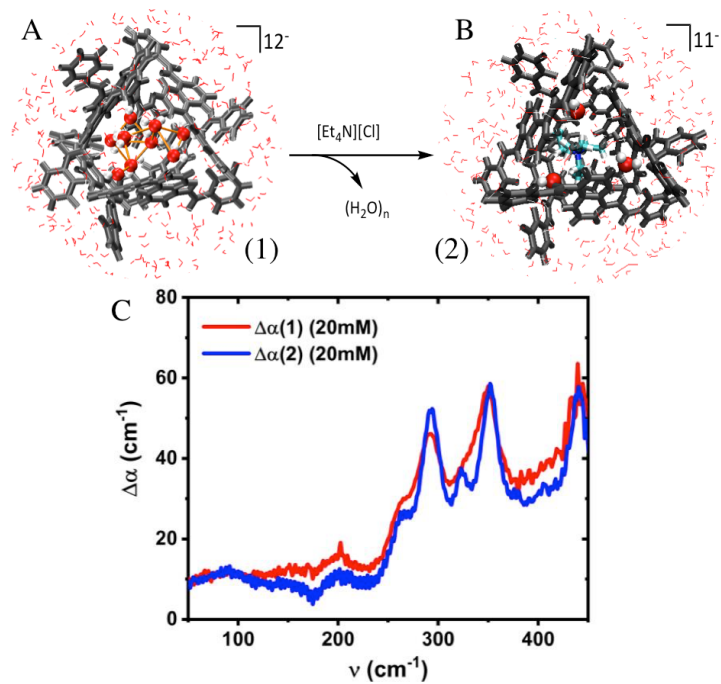

**Figure 1.** *THz spectroscopy performed on the Ga$_4$L$_6^{12-}$ tetrahedron with water vs [Et$_4$N]$^+$ salt guests in the internal cavity.* (a) Water molecules (oxygens in red, hydrogens in white) and the (b) cationic guest [Et$_4$N]$^+$ within the Ga$_4$L$_6^{12-}$ tetrahedron. The hydrogen bonds formed in between water molecules inside the cage (orange lines) are also shown, as well as the three water molecules remaining within the cage in presence of the guest. In the Ga$_4$L$_6^{12-}$ assembly the metal ions occupy the four vertices and the ligands (L) are bridging aromatic spacers, occupying each of the six edges of the tetrahedron (grey bonds), and have a length of 12.9 Å. (c) $\Delta\alpha(\nu)$ for water-filled (red line) and 20 mM [Et$_4$N]$^+$ guest filled (blue line) inside the nanocage after bulk water subtraction. All the absorption spectra were recorded under identical conditions (temperature, air humidity and concentration). Details of the THz setup (Figure S1) and difference spectra are provided in the Supporting Information.

Figure 2 shows a double difference, $\Delta\Delta\alpha(\nu)$, between the absorption of the guest-host complex in the presence ($\Delta\alpha(2)(\nu)$) vs. absence ($\Delta\alpha(1)(\nu)$) of the [Et$_4$N]$^+$ guest molecule to isolate the THz fingerprint of the water cluster in the cavity

$$\Delta\Delta\alpha(\nu)=\Delta\alpha(1)(\nu)-\Delta\alpha(2)(\nu) \qquad (1)$$

When the signal is normalized with respect to the host concentration, it is found that the intensity of $\Delta\Delta\alpha(\nu)$ is independent of the host concentration of 10mM or 20 mM. This provides validation that in these experiments that the number of water molecules inside the nanocage does not depend on the concentration of the supramolecular host, and indicates there is no aggregation or precipitation at the higher concentration.

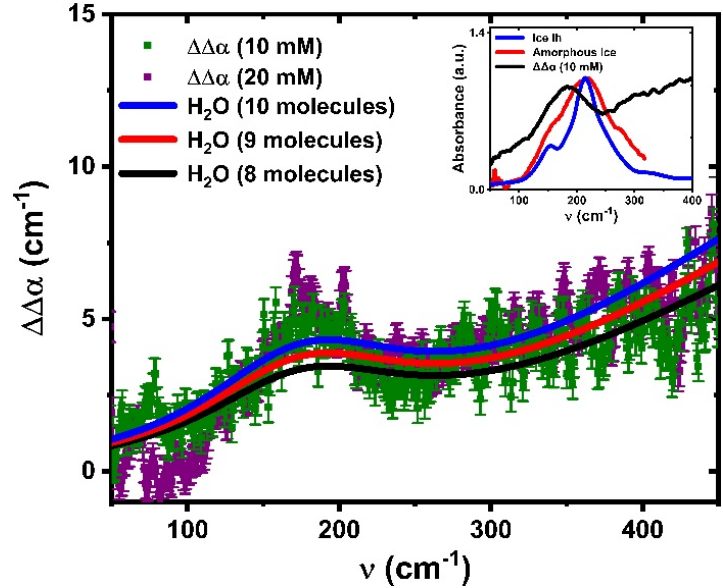

**Figure 2.** *Double difference spectrum, $\Delta\Delta\alpha(\nu)$, to isolate the THz fingerprint of the water cluster in the cavity of $Ga_4L_6^{12-}$.* Experimental data at 10mM (green points) and 20mM (purple points) solutions, at 293 K. The data at 10mM have been rescaled to the 20mM concentration for comparison. Inset: Absorbance spectra of ice Ih (blue line) from Ref.[38], low-density amorphous ice (red line) from Ref.[35] and $\Delta\Delta\alpha(\nu)$ at 10mM. The maximum absorption of each spectrum has been normalized to unity for the sake of comparison. The error of the absorption coefficients of bulk water, hexagonal ice and amorphous ice is less than 5%.

The $\Delta\Delta\alpha(\nu)$ intensity is then compared to a spectrum of bulk water that has been scaled by a number density to isolate the water count inside the cage. Inspection of Figure 2 indicates that when compared against the bulk spectrum scaled by 8, 9, or 10 water molecules (black, red and blue lines, respectively), it is proposed that 9±1 $H_2O$ molecules are dynamically confined inside the cage in the absence of the salt (also see Figure S2). This is in agreement with the estimate for the number of water molecules which can be hosted in the cavity, taking into account a total volume of 270 $Å^3$.[5] Even so, the estimated number of waters has to be considered as an average, resulting from the exchange of water molecules near the host interface with the bulk solvent through the open faces of the cage.

Not surprisingly, the spectrum of the encapsulated water shown in Figure 2 is very different from the sharp absorption features observed for gas phase water clusters.[44] Although the THz absorption spectroscopy is not a direct probe of the structure of a system, these low-frequency spectral signatures are specific fingerprints of the hydrogen bond network of the isolated water cluster in the $[Ga_4L_6]^{12-}$ host compared to other water systems.[34, 38] In previous structural studies on supramolecular hosts in the solid state, encapsulated hydrogen-bonded $(H_2O)_{8-10}$ water clusters were identified to be similar to the smallest subunit of cubic ice (Ic).[26-28] Even after soaking a supramolecular host crystal in water for few hours, a crystallized water decamer was observed in the cavity, albeit without a perfectly close-packed arrangement as in ice.[45] However, all these previous studies were of crystals and are not directly comparable to those carried out in solution under ambient conditions where the $[Ga_4L_6]^{12-}$ capsule can perform catalytic reactions.[6, 8-9]

Thus, to better determine the nature of the encapsulated water, its spectrum was compared to those of hexagonal or amorphous ice (Figure 2, Inset). The spectroscopic fingerprint of the confined water network lacks the characteristic peak at about 220 $cm^{-1}$ with a shoulder at 150 $cm^{-1}$ as observed in the case of Ic and hexagonal ice (Ih).[33, 38] In addition, the maximum of the band of water trapped inside the capsule is strongly redshifted with respect to the broad mode of low-density amorphous ice (at about 215 $cm^{-1}$), indicating a weaker hydrogen bond than the solid.[35] Finally, the librational band of liquid water, i.e. the increased intensity above 250 $cm^{-1}$, is clearly visible in $\Delta\Delta\alpha(\nu)$, while it is missing in ice at these frequencies. Thus the spectrum of the encapsulated water does not resemble the spectrum of amorphous ice nor that of Ih or Ic.

To compare the similarity of capsule-confined water to other water phases, we performed a detailed analysis of the center frequencies, $\tilde{\nu}_0$, which indicate the strength of the hydrogen bonds involved in the vibration, and linewidths, $w_0$, that yield information on the hydrogen bond network with regards to lifetime and the number of available chemical environments (i.e. the degrees of freedom).[46] Therefore the dynamics of the encapsulated water network, embodied in $\Delta\Delta\alpha(\nu)$ was fitted to a sum of three damped harmonic oscillators, describing the relaxational, the intermolecular hydrogen bond stretching, and the librational modes with increasing frequency.[36] The resulting decomposed spectrum is shown in Figure 3.

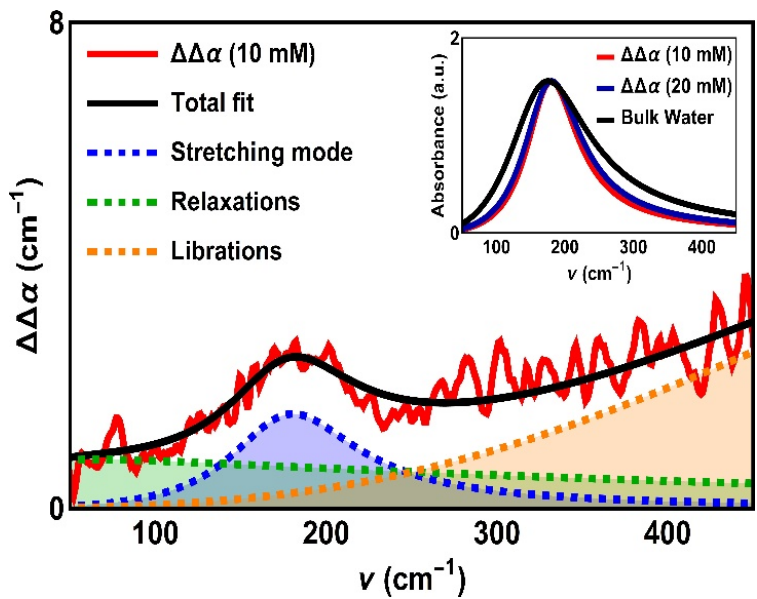

**Figure 3.** *THz difference spectrum, $\Delta\Delta\alpha(\nu)$, and its fit for a 10mM supramolecular host solution.* The different modes of the hydrogen bonding network are dissected with a fit for intermolecular relaxation, hydrogen bond stretching, and water librational modes (see text and Supporting Information for details). Inset: Fit of the intermolecular stretching band of $\Delta\Delta\alpha$ at 10 mM and 20 mM respectively, and of bulk water at room temperature. All the intensities have been rescaled to the maximum absorption of $\Delta\Delta\alpha$ at 10 mM.

The broad background extending to low-frequencies (<100 cm$^{-1}$) is attributed to dielectric relaxations and is found to be very similar to bulk water. The maximum of the librational peak (i.e. the hindered rotations) lies outside our experimental frequency range which stops at 450 cm$^{-1}$. Thus for the fits reported in Figure 3 the center frequency of the librational modes of the water confined in the cavity was fixed to 650 cm$^{-1}$ as in bulk water.[36] By closer inspection, the increase in absorption with increasing frequency from 180 to 400 cm$^{-1}$ is smaller than in the case of bulk water: $\Delta\Delta\alpha(400\text{ cm}^{-1})/\Delta\Delta\alpha(180\text{ cm}^{-1})=1.02\text{-}1.10$ for confined water while $\Delta\Delta\alpha(400\text{ cm}^{-1})/\Delta\Delta\alpha(180\text{ cm}^{-1})=1.45$ for bulk water. This is indicative of a blue shift of the librational mode, which can be attributed to a strong steric hindrance encountered by the librations of the water molecules in the proximity of the cage's internal surface. A similar linewidth narrowing of the librational mode was found for water confined in nanoporous silica glasses, but in that case it exhibited a blueshift of the peak frequency itself due to interaction with the hydrophilic matrix.[47]

The most interesting part of the THz spectra arises from the observation of an unperturbed center frequency of the intermolecular hydrogen bond stretching mode of water confined in the $Ga_4L_6^{12-}$ cage. Table 1 provides the values of $\tilde{\nu}_0$ and $w_0$ in which the stretch band is centered at 180 cm$^{-1}$ for confined water which is (perhaps surprisingly) not shifted with respect to the center frequency of bulk water at 293 K (181 cm$^{-1}$). The intermolecular vibrations of the confined water are clearly redshifted by ~10 cm$^{-1}$ with respect to the same mode for water cooled to its freezing point[38], and by ~35 cm$^{-1}$ with respect to water under high (~10 kbar) hydrostatic pressures (see Table 1 and Figure S3).[37] Thus, the nanoconfined water cannot be considered as cold or pressurized water either, but indicates a similar intermolecular hydrogen bond strength like that of ambient water.

At the same time, the confined water shows a significant decrease in the damping of the intermolecular stretching mode, characterized as a significant narrowing of the linewidth with respect to bulk water at 293 K (Table 1). Any decrease in linewidth is an indicator for a decreased variance in the fast dynamics[36] (see the Supporting Information), and has also been ascribable to a reduced number of degrees of freedom, i.e. an entropic signature of a more restricted set of molecular configurations that are available.[46] To place the linewidth of the nanoconfined water into perspective, we find that its value of $w_0$ =250 cm$^{-1}$ is greatly reduced with respect to ambient, cold, and pressurized bulk water (~540 cm$^{-1}$), as well as with respect to the two hydration bands around the hydrophobic groups of alcohol chains and lightly supercooled water at 266.6 K that exhibit linewidths between 340-440 cm$^{-1}$. Instead, the observed linewidth of the confined water interpolates between that observed for hexagonal ice ($w_0$ =80-220 cm$^{-1}$ and clathrate hydrates and amorphous ice ($w_0$ =280-300 cm$^{-1}$).[37-38]

**Table 1.** Spectral parameters of the intermolecular stretching band of water confined in the nanocapsule and bulk water at different thermodynamic conditions. Parameters are obtained by fitting a set of damped harmonic oscillators. The statistical 2σ error is given in brackets. Further details and the results of the fit can be found in Tables S1 and S2 and Figures S3 in the Supporting Information. (also see Refs.[38] and [37] for further details)

| Fit Parameter (cm$^{-1}$) | Water inside Ga$_4$L$_6$ | Water (293 K) | Water (273.2 K) | Water (10 kbar) |
|---|---|---|---|---|
| $\tilde{\nu}_0$ | 180 (4) | 181 (2) | 193 (2) | 216 (4) |
| $w_0$ | 249(18) | 537 (3) | 557 (4) | 542 (9) |

*Theoretical Results.* To provide support for the experimental interpretations of the dynamics and structure described above, we have performed AIMD simulations of the solvated [Ga$_4$L$_6$]$^{12-}$ host to characterize the encapsulated water molecules, using a well-characterized meta-GGA functional B97M-rV[48], shown to describe bulk water well.[49-50] Figure 4a provides the AIMD simulated THz spectra of water inside the cage and the bulk water spectrum compared to experiment (see Methods). The theoretical spectrum reproduces accurately the two main features of the THz measurements: (1) the same position of the intermolecular hydrogen-bonded stretching band at 180 cm$^{-1}$ for both water inside the cage and in the bulk, and (2) the reduction in linewidth for water inside the cage with respect to bulk water. An AIMD additional simulation with the [Et$_4$N]$^+$ guest does not exhibit differences in interfacial properties near the cage (Figure S4), and thus does not contribute to the difference THz spectra.

Given the excellent agreement, we analyzed the trajectories and determined that the time-averaged number of water oxygen centers inside the salt-free cage is 12.4 ± 0.7, whereas we find an average of 3.4±0.6 water molecules inside the cage when filled with the cationic substrate. However, in the water-filled cage there are 9 water molecules that are dynamically distinct, with long residence times within the cage on timescales that exceed that of the 1-3 ps timescale of bulk water[51-52] by at least an order of magnitude, whereas the remaining ~3-4 waters undergo fast exchange dynamics with the bulk water at the interface like that of traditional guest-filled substrates (see Figure S4).

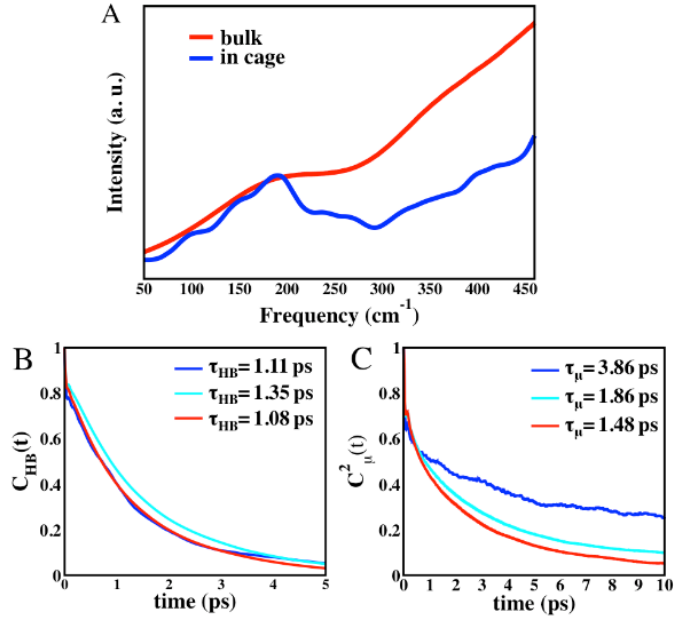

**Figure 4.** *Water hydrogen bond dynamics inside and outside the $Ga_4L_6^{12-}$ cage.* (a) Theoretical THz-IR spectra calculated for water inside the cage (blue) and for bulk liquid water taken from previous work[49] (red). The intensities are rescaled in order to have the same intensity for the maximum at ~180 cm$^{-1}$ to aid comparison. (b) hydrogen-bonded lifetimes $C_{HB}(t)$ and (c) orientation correlation $C_\mu^{(2)}(t)$ dynamics inside the cage (blue), in the hydration layer (cyan) and in the bulk liquid (red). The characteristic relaxation times are also reported in the legend. Details on defined regions are given in Figure S5 and text in the Supplementary Information.

To quantify these motions for the water filled $Ga_4L_6^{12-}$ cage, we evaluate the lifetime of hydrogen-bonds (HB) using the intermittent water-water HB autocorrelation function, $C_{HB}(t)$

$$C_{HB}(t) = \frac{\langle h(t)h(0)\rangle}{\langle h(0)\rangle^2} \qquad (2)$$

where the operator h(t) is 1 when a given HB is intact and 0 otherwise.[53] We find that the HB-lifetime $\tau_{HB}$ is very similar for all water regions ~1 ps (Figure 4b). This is in agreement with the experimental observation that the central frequency of the 180 cm$^{-1}$ band is unshifted for water inside of the cage and the bulk, and an indicator of similar HB energetics.[54] We have also calculated the orientational correlation function of the dipole vector of the water molecules as:

$$C_\mu^{(2)}(t) = \frac{\langle P_2[\mu(t)\cdot\mu(0)]\rangle}{\langle P_2[\mu(0)\cdot\mu(0)]\rangle} \qquad (3)$$

where $P_2$ is the 2nd rank Legendre polynomial and $\mu(t)$ is the water dipole moment (unit vector) at time t.[55-56] Inspection of Figure 4c reveals that water orientational dynamics is remarkably slower inside the cage when compared to the hydration and bulk water, in which the orientation relaxation time, $\tau_\mu^{(2)}$ is ~2.5 times longer for water molecules in the host. The slowdown of water orientational dynamics inside the cage can be rationalized in terms of the constraints imposed by the confinement on allowed reorientations and thus fewer hydrogen-bonded network configurations available within the cage with respect to the external solvent. This result is in agreement with the speculations made on the experimental side in regards the linewidth analysis of the spectral band at 180 cm$^{-1}$, which is found to be much sharper for water inside the cage than for bulk liquid water. This signature is likely attributable to the lost translational motion as measured by long residence times as well as arising from restricted rotational motions for water in the cage.

To characterize the "phase" of water within the cage, as much as can be said for such a small cluster, we consider two popular structural order parameters used to describe the structure, dynamics, and thermodynamics of bulk water over its phase diagram.[57] The translational order parameters, t, is defined as

$$t = \frac{\int_0^{\xi_c} d\xi \, |g_{OO}(\xi)-1|^2}{\xi_c} \qquad (4)$$

where $g_{OO}(\xi)$ is the oxygen-oxygen radial distribution function, $\xi = r\rho^{1/3}$, r is the distance between the oxygen atoms of a pair of molecules, $\rho$ is the bulk water density, and $\xi_c$ is a cut-off distance that we set to 3Å in this work. For an ideal gas, $g(\xi) = 1$ everywhere and t vanishes, whereas in a crystal there is long-range translational order, and $g(\xi) \neq 1$ over long distances and hence t is large. For example, values of t = 1 can be obtained for an fcc crystal such as cubic ice.[58] The *q* parameter

$$q = 1 - \frac{3}{8} \sum_{i=1}^{3} \sum_{j=i+1}^{4} \left(cos\phi_{ij} - \frac{1}{3}\right)^2 \qquad (5)$$

measures tetrahedral order, where $cos\phi_{ij}$ is the angle formed by the lines joining the oxygen atom of a given molecule and those of its nearest i,j neighbours ($\leq 4$). The average value of *q* varies between 0 (in an ideal gas) and 1 (in a perfect tetrahedral network, as it is the case for ice).

Figure 5a shows that water within the cage is different from the interfacial water near the nanocage interface, bulk (liquid) water, and ice. When considering the order parameters for water inside the cage, one can in particular notice that $t$ is larger than for bulk liquid water, while the opposite is true for $q$. This $t-q$ trend has been shown previously to occur when bulk water is isothermally compressed at a low temperature.[57] From this, one would be tempted to correlate the phase of water inside the cage to that of pressurized water.

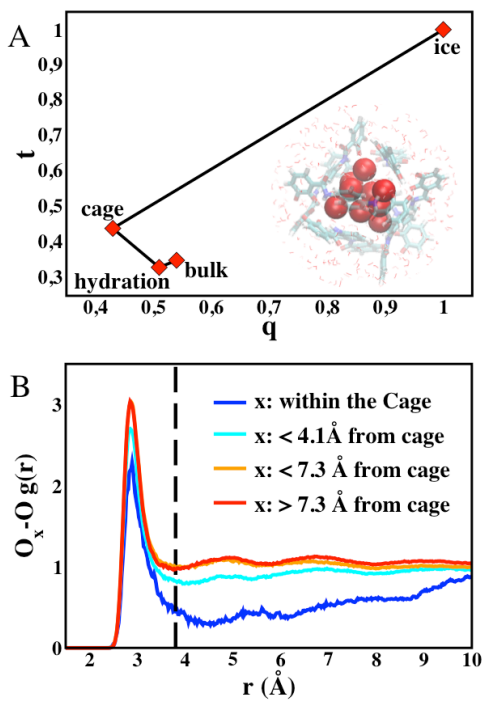

**Figure 5.** *Structural characterization of the nanoconfined water droplet.* (a) The $t$ and $q$ order parameter space showing their values for cubic ice, water confined inside the cage, water in the hydration layer outside the cage (<4.1 Å), and bulk liquid water. The $t$ parameter value for cubic ice (t=1) is estimated for a fcc crystal[58], while that the q parameter (q=1) is that of a perfect tetrahedral environment. The other $q$ and $t$ values can be found in Table 2. The inset illustrates the average position of the 9 arrested water molecules with long residence time inside the cage. (b) Oxygen-oxygen radial distribution function $g_{OO}(r)$ for water inside the cage (blue) compared to the bulk (red).

This assumption can be checked by analyzing the $g_{OO}(r)$ for water inside the cage and in the bulk as shown in Figure 5b. In recent work it has been shown that when the pressure is increased on the water liquid, the 4-5 waters residing in the region of the first peak of $g_{OO}(r)$ are nearly unchanged, whereas in the region beyond the first peak large structural changes occur with the collapse of the 2nd hydration shell and shifting of higher shells to shorter distances.[59] On the contrary we find that $g_{OO}(r)$ for water in the cage shows a less intense peak with respect to bulk water, with no significant density in the outer shells. From this we can infer that water within the cage is not equivalent to pressurized water, despite the fact that they have some similarities in terms of $t-q$ order parameters. When the same aqueous $Ga_4L_6^{12-}$ supramolecular tetrahedral assembly is simulated at a low temperature of 260K, the conclusion that the water inside the cage is remarkably different from bulk water and ice does not change (see Figure S5).

Table 2 summarizes the $t$ and $q$ order parameters for cubic ice, water confined inside the cage, water in the hydration layer outside the cage (<4.1 Å), and bulk liquid water. In Table 2, we also report the water coordination number as defined by integration under the first peak of the $g_{00}(r)$ at various cutoff values, as well as the number of HBs per molecule calculated using the distance and angle criterion prescribed by Luzar.[53] All the structural signatures in Table 2 suggest that water molecules inside the cage are severely undercoordinated, with greatly reduced hydrogen-bonding with respect to bulk liquid water. In particular, water inside the cage forms on average 1.8 HBs/molecule compared to 3.4 HBs/molecule formed in bulk liquid water and 4 HBs/molecule formed in a perfect tetrahedral ice structure. Among the 1.8 HBs/molecule formed by water inside the capsule, an average of 1.5 HBs/molecule is formed between waters inside the cage, while only 0.3 HBs/molecule are formed with water at the cage interface.

**Table 2.** Hydrogen bond index per molecule, coordination number (CN) with different cutoff values of the first shell, orientational (q) and translational (t) order parameters for bulk water, interfacial water at the cage surface and water inside the cage.

| Water System | HBs per molecule | CN (3.2Å) | CN (3.8Å) | q | t |
|---|---|---|---|---|---|
| bulk | 3.4 | 3.8 | 6.6 | 0.54 | 0.35 |
| hydration | 2.9 | 3.4 | 5.8 | 0.50 | 0.33 |
| cage | 1.8 | 2.8 | 4.5 | 0.43 | 0.44 |

The water undercoordination suggests that, instead of an ice-like structure, water within the cage most likely behaves as an isolated small droplet with fixed structure unlike other bulk phases. This is in agreement with the conclusion reached from the THz dynamical experiment, which finds that the encapsulated water does not resemble any testable phase of water, instead exhibiting mixed spectroscopic signatures of the liquid and solid phases over different parts of the water phase diagram.

**DISCUSSION AND CONCLUSION**

In summary, this work is the first experimental and theoretical characterization of the structure and dynamics of the water confined in the $Ga_4L_6^{12-}$ tetrahedral assembly in solution under ambient conditions. Although we did not observe a shift of the intermolecular hydrogen bond stretching center frequency with respect to bulk water at room temperature, indicating that the bond strength and length are not affected by confinement, the linewidth of this band is found to be about 55% smaller than in bulk water and more similar to that of amorphous ice or ice clathrates. This particular feature in the spectrum is a direct signature of the reduced number of degrees of freedom of the water molecules in confinement caused by a reduced number of available hydrogen bonds to create a collective hydrogen-bonded network. Moreover, the librational motions of water, which are facilitated in a three-dimensional network since they involve a cooperative motion, as in the well-known jump mechanism[60], are restricted by the steric hindrance near the hydrophobic surface of the cavity.

The integrated results indicate that water confined in the $Ga_4L_6^{12-}$ supramolecular host is not similar to water in any other thermodynamic state (e.g. at low temperature and/or high pressure), as also recently suggested by Heyden and co-workers for proteins[61]. Supporting AIMD simulations show that the dynamical signatures of the water droplet indicate that it is strongly arrested, and that it has a disrupted hydrogen bond network on its outer layer, with hydrogen bonding maintained within the core of the water droplet. The simulations also support the spectroscopic interpretation of the narrowing of the linewidth of the intermolecular stretching mode due to reduced translational and rotational motions of the confined water.

This implies that any release of water from the host cavity into the bulk will be entropically favourable. The release of the cavity waters is also enthalpically favoured because the confined water cannot form as many hydrogen bonds as in the bulk, and thus are "high energy" or "frustrated".[62-64] However, as always there is an enthalpy-entropy balance that must be considered in the overall desolvation process, requiring the stripping of water molecules from the reactant and subsequent preferential solvation of the transition state.[65] But fundamentally, the soluble $Ga_4L_6^{12-}$ cage does create an inherent thermodynamic drive for guest encapsulation through desolvation of the host cavity.[11-14, 62]

**MATERIALS AND METHODS**

The Ga-host synthesis has been reported previously.[66-67] The 1:1 binding with $Et_4N^+$ is verified by $^1$H-NMR of the synthesized host, which shows encapsulated $Et_4N^+$ and no free salt in solution. See supplementary material for details about the sample preparation, measurements and data analysis.

*THz spectroscopy.* Spectra of Gallium supramolecular hosts aqueous solutions at 10 and 20 mM were recorded at 293 K in the frequency range from 50 to 450 cm$^{-1}$ by THz-Far Infrared (THz) absorption spectroscopy. THz measurements were performed using a Bruker Vertex 80v FTIR spectrometer equipped with a liquid helium-cooled bolometer from Infrared Laboratories as detector. The sample solutions were placed in a temperature-controlled liquid transmission cell with polycrystalline diamond windows and a 25 μm-thick Kapton spacer. 128 scans with a resolution of 2 cm$^{-1}$ were averaged for each spectrum. The double difference absorption spectra were smoothened with a 2 cm$^{-1}$ wide (5 point) moving average.

*Ab initio molecular dynamics.* All calculations presented in this paper were performed with Density Functional Theory (DFT) using the dispersion corrected meta-generalized gradient approximation (meta-GGA) functional B97M-rV[48, 68-69] in combination with a DZVP basis set optimized for multigrid integration[70] as implemented in the CP2K software package.[71-72] In all cases, we used periodic boundary conditions, 5 grids and a cutoff of 400 Ry. Three independent AIMD simulations were performed for 30 ps in the NVE ensemble after an equilibration period of 6 ps (3ps in the NVT ensemble with T=300K followed by 3ps in the NVE ensemble). In the NVE trajectories, the average temperature was 318±9 K. All results are based on averages over the three AIMD simulations. The time-averaged number of water inside the cage has been defined for each of the 3 independent simulations by counting at each step the number of waters within the cage and averaging over all MD steps.

*THz Spectra Simulation.* The theoretical IR spectra in the 50-500 cm$^{-1}$ THz frequency range were calculated using the strategy developed recently based on the Fourier transform of the velocity-velocity correlation function modulated by Atomic Polar Tensors (APT)[73-74]:

$$I(\omega) = \frac{2\pi\beta}{3cV} \sum_{u=x,y,z} \sum_{m=1}^{3N} \sum_{l=1}^{3N} \int_{-\infty}^{+\infty} dt\, e^{i\omega t} \langle P_{um}(t)v_m(t)P_{ul}(0)v_l(0)\rangle \quad (6)$$

where $\beta = 1/kT$, $\omega$ is the frequency, c the speed of light, V the volume of the system, $\langle \cdots \rangle$ the equilibrium time correlation function, N the number of atoms of the system, $v_m$ the m$^{th}$ element of the $\boldsymbol{v}$ vector that collects the 3N cartesian velocities of the N atoms of the system. $P_{um} = \partial\mu_u/\partial\xi_m$ is the um element of the atomic polar tensor, i.e. the first derivative of the u$^{th}$ component (u=x,y,z) of the total dipole moment $\boldsymbol{M}$ of the system with respect to the m$^{th}$ cartesian coordinate. The above equation takes into account all self- and cross-correlation terms, whether intra- or inter-molecular, as well as both the charges and the charge fluxes contributions to the IR intensity, and simultaneously reduces the computational cost from the usual Fourier transform of the dipole moment correlation and accelerates signal convergence, without loss in accuracy.[74] Velocities ($v_m$) are readily obtained from the DFT-MD trajectories while $\boldsymbol{P}(t)$ tensors have been parameterized on reference water structures.[73] The spectra are calculated including water contributions only, and neglecting contribution from the cage and counter-ions. By selecting the cartesian coordinates of the atoms belonging to a specific vibrational population (class of water molecules with common structural and spectroscopic properties) into the summation in eq. (4), one gets the individual contribution of the selected population to the IR spectrum by reducing the summation over m=1,3N to m=1,3N*, where N* = 9 identifies the waters inside the cage.

**SUPPORTING INFORMATION**

The Supporting Information is available free of charge on the PNAS Publications website. Included in the Supporting Information are experimental details, characterization data, and methods available as a PDF.

**ACKNOWLEDGMENTS**

This work was funded by the Deutsche Forschungsgemeinschaft (DFG, German Research Foundation) under Germany´s Excellence Strategy – EXC 2033 – 390677874 – RESOLV, and by the DFG Research Training Group (GRK 2376/331085229) "Confinement-controlled Chemistry". MH acknowledges funding from the ERC Advanced Grant 695437 THz calorimetry. The Berkeley scientists are supported by the Director, Office of Science, Office of Basic Energy Sciences, and the Division of Chemical Sciences, Geosciences, and Bioscience of the U.S. Department of Energy at Lawrence Berkeley National Laboratory (Grant No. DE-AC02-05CH11231) and a NIH Postdoctoral Fellowship to T.A.B. (Grant No. 1F32GM129933-01). This research used resources of the National Energy Research Scientific Computing Center, a DOE Office of Science User Facility supported by the Office of Science of the U.S. Department of Energy under Contract No. DE-AC02-05CH11231. We also want to thank Dr. Daria R. Galimberti and Prof. Marie-Pierre Gaigeot for sharing the codes for the THz spectra calculations.